\documentclass[twoside,slac_one]{revtex4}
\usepackage{graphicx}
\usepackage{fancyhdr}
\usepackage{amsmath} 
\usepackage{bm}
\usepackage{amsxtra}
\usepackage{amssymb}
\usepackage{amsthm}
\usepackage{latexsym}
\usepackage{lscape}
\usepackage{xspace}
\usepackage{subfigure}

\def\MjjLowerLimitMRST{1265} 
\def\MjjLowerLimitMRSTEXPECTED{1100} 
\def\MjjLowerLimitMRST{2.15} 
\def\MjjLowerLimitMRSTEXPECTED{2.07} 
\def\MjjLowerLimitMRSTAxigluon{2.10} 
\def\MjjLowerLimitMRSTEXPECTEDAxigluon{2.01} 

\def\MjjLowerLimitMRSTQBH{3.67} 
\def\MjjLowerLimitMRSTEXPECTEDQBH{3.64} 
\def\FchimjjLambda{9.5}  
\def\FchimjjEXPECTEDLambda{5.7}  
\def\FchiKFacQCDCILambda{6.8}  
\def\FchiKFacQCDCIEXPECTEDLambda{5.2}  
\def\FchimjjQstar{2.64}
\def\FchimjjEXPECTEDQstar{2.12}
\def\FchimjjQBH{3.78}
\def\FchimjjEXPECTEDQBH{3.49}

\def\FchiLowerLimitMRSTQBHnsix{3.69}
\def\FchiLowerLimitMRSTEXPECTEDQBHnsix{3.37}

\def\ElevenBinChiEXPECTEDLambda{5.4}
\def\ElevenBinChiLambda{6.7}
\def\ElevenBinChiEXPECTEDQBH{3.46}
\def\ElevenBinChiQBH{3.49}

 \def\mFchi{F_{\chi}}
 \def\mFchimjj{F_{\chi}(m_{jj})}
 \def\Fchi{$\mFchi$}
 \def\Fchimjj{$\mFchimjj$}

\def\ipb{pb$^{-1}$\xspace}
\def\ifb{fb$^{-1}$\xspace}
\def\mjj{\ensuremath{m_{jj}}\xspace}
\def\pt{$p_T$\xspace}
\def\pval{\ensuremath{p\mbox{-value}}\xspace}
\def\pvals{$p$-values\xspace}

\def\lumi{1.0 fb$^{-1}$\xspace}
\def\lumUncert{3.7}
\def\figdir{figures}
\def\bhpval{82}
\def\pvalL{28}
\def\limExqExp{2.81}
\def\limExqObs{2.99}
\def\limAExp{3.07}
\def\limAObs{3.32}
\def\limSExp{1.77}
\def\limSObs{1.92}

\def\lumi{0.81 fb$^{-1}$\xspace}
\def\lumUncert{4.5}
\def\figdir{figures081}
\def\bhpval{62}
\def\pvalL{13}
\def\limExqExp{2.77}
\def\limExqObs{2.91}
\def\limAExp{3.02}
\def\limAObs{3.21}
\def\limSExp{1.71}
\def\limSObs{1.91}

\begin{document}

\title{Dijet searches for new physics in ATLAS}

%

\author{Georgios Choudalakis}
\affiliation{Enrico Fermi Institute, University of Chicago, Chicago, IL, USA}

\begin{abstract}
The latest results are presented of the search for new physics in inclusive dijet events recorded with the ATLAS detector.  The search for resonances in the dijet mass spectrum is updated with \lumi of 2011 data.  The latest analysis of dijet angular distributions, with 36 \ipb of 2010 data, is also presented.  In-depth information is provided about the model-independent search for resonances.  Limits are provided for excited quarks, axigluons, scalar color octets, and to Gaussian signals that allow to set approximate limits in a model-independent way.
\end{abstract}

\maketitle

\thispagestyle{fancy}

\section{Introduction}

A search for new physics in events with at least two hadronic jets is motivated by numerous proposed extensions of the Standard Model (SM), ranging from quark structure to extra spatial dimensions.  Experimentally, the large number of dijet events allows for a data-driven background estimation from very early\footnote{Indicatively, \cite{ICHEP,PRL} was the first LHC search for exotic phenomena to reach beyond Tevatron limits, with only 0.3 \ipb.}.

ATLAS currently pursues two complementary searches in dijet events: 
\begin{enumerate}
\item The search for resonances in the mass (\mjj) spectrum of the two highest-\pt (a.k.a.\ ``leading'') jets, \cite{PRL, NJP}
\item The search for an enhancement of leading jets produced at similar rapidities, i.e.\ at small $|\Delta y|$. \cite{angularPLB, NJP}
\end{enumerate}
For brevity, the former analysis is called ``Resonance Search'', and the latter ``Angular Search''. Both are sensitive to very similar new physics that would appear in both \mjj and $\Delta y$.

Section \ref{sec:flashback} offers an overview of the Angular Search and its latest results with 36 \ipb of 2010 data.  Section \ref{sec:newResults} contains the latest update of the Resonance Search, with \lumi of 2011 data.  In Section \ref{sec:searchPhase}, the opportunity is taken to offer in-depth information about the method used to search for anomalies in a model-independent way, explaining hypertests and the {\sc BumpHunter} \cite{BH}.  Section~\ref{sec:limits} summarizes the limits set to specific models, and Sec.~\ref{sec:MIlimits} the model-independent limits.

\section{Summary of 2010 results}
\label{sec:flashback}

This Section summarizes the results of \cite{NJP}, where both dijet searches were presented with 36 \ipb of 2010 data.  Since this has been the latest update of the Angular Search, more emphasis will be given to it here, deferring the Resonance Search for later.

\begin{figure}
\subfigure[The observed (filled points) and expected (histograms) $\chi$ distribution.]{
\includegraphics[width=0.45\textwidth]{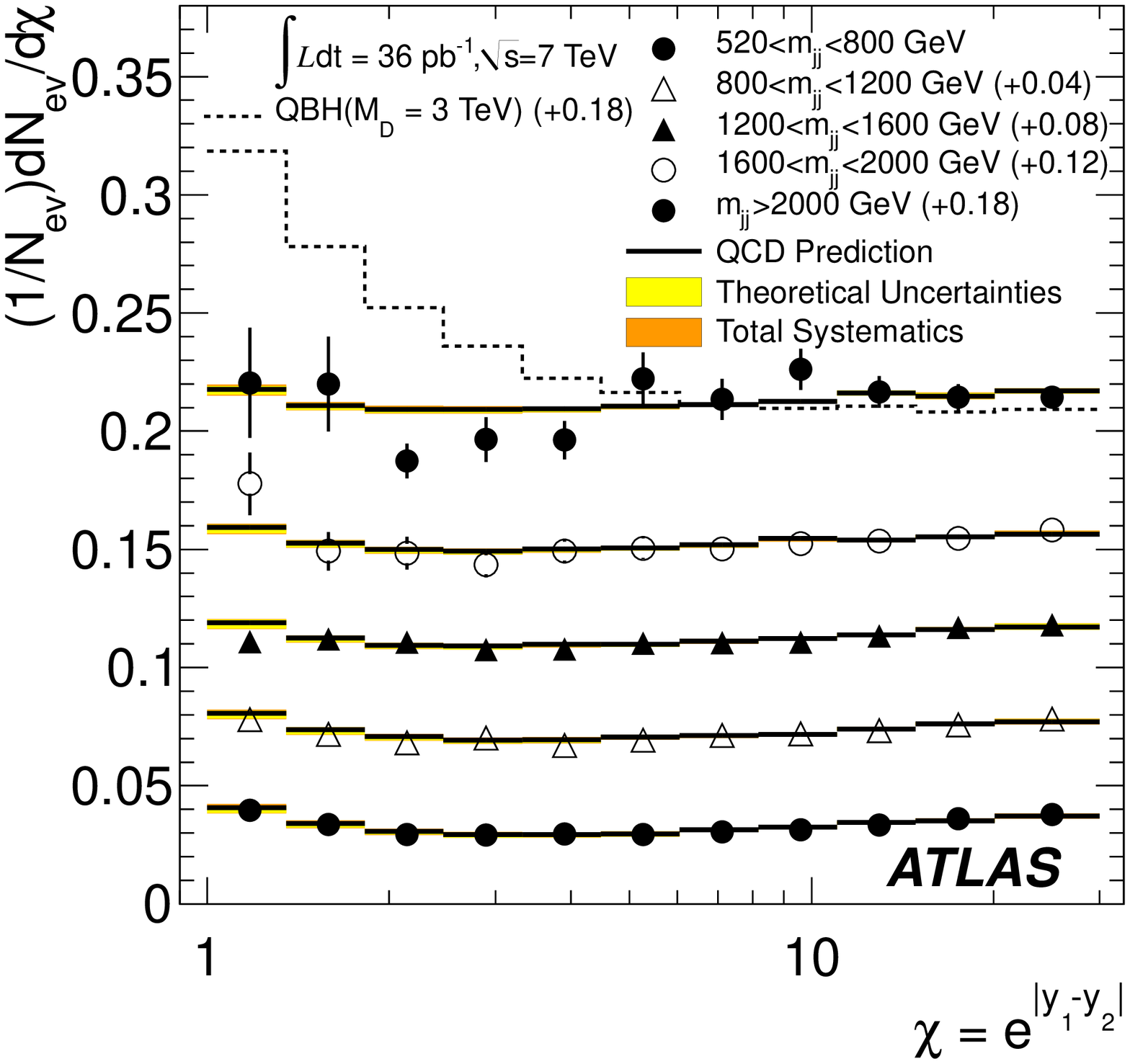}
\label{fig:chi}
}
\subfigure[The observed (filled points) and expected (histograms) $F_\chi(\mjj)$ distribution.]{
\includegraphics[width=0.45\textwidth]{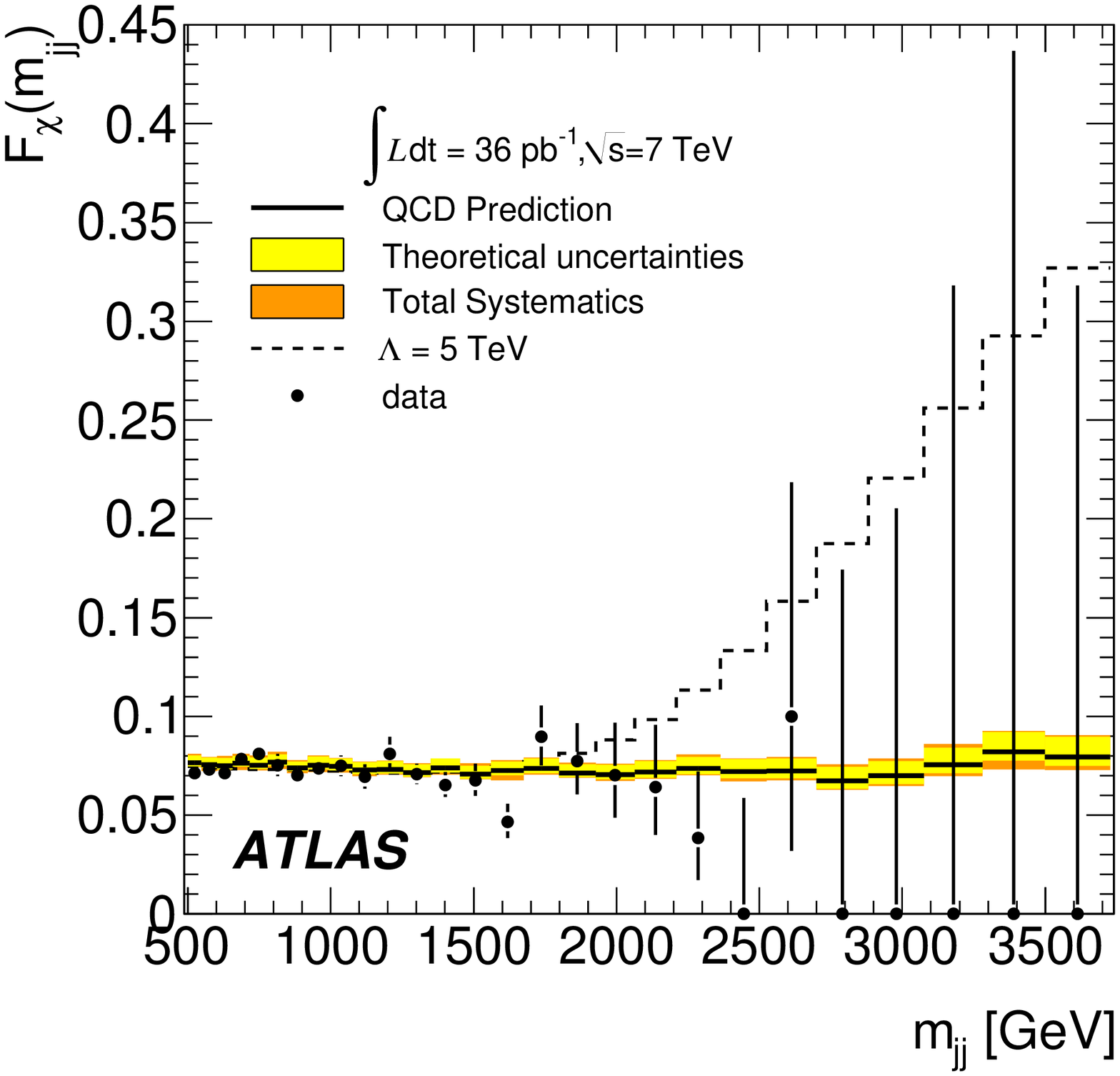}
\label{fig:Fchimjj}
}
\caption{Distributions of $\chi$ (left) and $F_\chi(\mjj)$ (right).  The dashed histogram is an example of quark compositeness signal MC. \label{fig:angularObservables}}
\end{figure}

The event selection, which is detailed in \cite{NJP}, ensures good data quality, well-measured jets, and constant trigger efficiency.  The basic observable of the Angular Search is
\begin{equation}
\chi \equiv e^{|\Delta y|},
\end{equation}
where $\Delta y$ is the rapidity\footnote{Rapidity ($y$) follows the usual definition with respect to the beam axis ($z$): $y \equiv \ln \frac{E+p_z}{E-p_z}$.} difference between the leading and the subleading jet, i.e.\ the jet with the highest and second highest \pt in each event.  Fig.\ \ref{fig:chi} shows the distribution of data in $\chi$, in 5 broad intervals (a.k.a.\ ``bins'') of $\mjj$.  The data are statistically compatible with the expectation, which is computed using {\sc Pythia} Monte Carlo (MC) to model QCD, and {\sc NLOJET++} for next-to-leading-order (NLO) corrections.
A derived observable is $F_\chi$, which is the fraction of events at $\chi < e^{1.2}$.  The choice of $e^{1.2}$ is based on optimization of the sensitivity to quark contact interactions.  New physics would appear as an in crease in $F_\chi$.  When $F_\chi$ is computed in bins of \mjj, the result is an spectrum which can indicate new physics by an increase of $F_\chi(\mjj)$ in \mjj bins that contain significant amounts of signal.  Fig.\ \ref{fig:Fchimjj} shows the observed and expected $F_\chi(\mjj)$ spectra.

No significant discrepancy is observed.  The \pvals of goodness-of-fit tests, which use as test statistic the negative logarithm of the likelihood ($L$) of the data in all bins assuming the background ($-\log L(\text{data}|\text{background})$), are of order 30\% or more.  This level of agreement is observed in all \mjj bins where $\chi$ was examined, and in the $F_\chi(\mjj)$ spectrum (Fig.~\ref{fig:angularObservables}).

Since agreement was found with the SM, limits were set to various new physics models.  Three angular observables were used:  The full \Fchimjj\ spectrum,  the $F_\chi$ of events with $\mjj > 2$~TeV, and the differential distribution of events in $\chi$, i.e.\ $\frac{dN}{d\chi}$, in events with $\mjj > 2$~TeV.   Table \ref{table:Limits} summarizes all results, including those of the Resonance Search, which have now been superseeded by 2011 data, as will be shown.   All Frequentist limits in \cite{NJP} apply the classical ($CL_{s+b}$) Neyman construction \cite{Neyman} with test statistic being the logarithm of the ratio of likelihoods of the hypothesis with and without signal.  Systematic uncertainties are mainly from the jet energy scale, the renormalization and factorization scales ($\mu_R$,$\mu_F$), and parton density functions (PDF).  These uncertainties are propagated to the limits by random sampling of the nuisance parameters, making thus the Neyman band wider, which is known to not be fully conistent with the Frequentist framework, but is a practice accepted as common.  Regarding in particular the limit on contact interaction scale $\Lambda$ using \Fchimjj, the observed Frequentist limit of \FchimjjLambda~TeV is much above the expected limit.   However this is not the case for the Bayesian limit using the same \Fchimjj, or for the Frequentist limits using the other two angular observables (see last 4 rows of Table \ref{table:Limits}). 

\begin{table}
\caption{Lower limits, at 95\% confidence  level (or credibility level, for Bayesian limits) (C.L.), set by the Resonance and Angular Search in 36 \ipb of 2010 data \cite{NJP}.  Results of the Angular Search are distinguished by the observable listed in the first column.   The second column lists which limits are Frequentist and which Bayesian (with a constant prior in signal cross-section for the Resonance Search, and in $\frac{1}{\Lambda^2}$ for the Angular Search).  \label{table:Limits}}
\begin{tabular}{ll|cc}
\hline
Analysis / observable  \quad \quad & Method & \multicolumn{2}{c}{95\%\ C.L. Limits (TeV)} \\
    &        &  Expected  &  Observed\\
      \hline

\multicolumn{4}{c}{\bf Excited Quark $\bf{q^\ast}$}  \\ \hline
Resonance in $m_{jj}$ & Bayesian &   \MjjLowerLimitMRSTEXPECTED & \MjjLowerLimitMRST     \\ 
\Fchimjj\ & Frequentist   &  { \FchimjjEXPECTEDQstar}  & { \FchimjjQstar} \\    

\hline \multicolumn{4}{c}{\bf Randall-Meade Quantum Black Hole for $\bf n=6$} \\ \hline
Resonance in $m_{jj}$ & Bayesian &  { \MjjLowerLimitMRSTEXPECTEDQBH} &  { \MjjLowerLimitMRSTQBH }   \\
\Fchimjj\  & Frequentist      &   \FchimjjEXPECTEDQBH  &  \FchimjjQBH  \\    
\Fchi\  for $\mjj > 2$~TeV & Frequentist  &   \FchiLowerLimitMRSTEXPECTEDQBHnsix & \FchiLowerLimitMRSTQBHnsix \\
$\frac{dN}{d\chi}$ for $\mjj > 2$~TeV & Frequentist  & \ElevenBinChiEXPECTEDQBH & \ElevenBinChiQBH \\

\hline \multicolumn{4}{c}{\bf Axigluon} \\ \hline
 Resonance in $\mjj$ & Bayesian  & { \MjjLowerLimitMRSTEXPECTEDAxigluon} & { \MjjLowerLimitMRSTAxigluon} \\  
\hline \multicolumn{4}{c}{\bf Contact Interaction $\bf \Lambda$} \\ \hline
\Fchimjj & Frequentist  &   { \FchimjjEXPECTEDLambda}  & {  \FchimjjLambda } \\
\Fchimjj & Bayesian &   { 5.7 }  & { 6.7 } \\
\Fchi\ for $\mjj > 2$~TeV & Frequentist  &  \FchiKFacQCDCIEXPECTEDLambda  &  \FchiKFacQCDCILambda  \\ 
$\frac{dN}{d\chi}$ for $\mjj > 2$~TeV & Frequentist & \ElevenBinChiEXPECTEDLambda & \ElevenBinChiLambda \\
\hline
\end{tabular}
\end{table}

\section{Dijet Resonance Search with \lumi of 2011 data}
\label{sec:newResults}

This section is a summary of \cite{resonancePLB}, that highlights some of the technical details of the analysis.  

The main observable is \mjj, the mass of the system of the two leading jets in \pt.  Events are selected to ensure good beam and detector conditions, well-reconstructed leading jets, no potential of accidental swapping of the order of jets in \pt, and constant trigger efficiency in \mjj.  To suppress the QCD background at high mass, the $|\Delta y|$ is required to be less than 1.2, in accordance with the Angular Analysis (Section~\ref{sec:flashback}).

The \mjj spectrum is compared to the expected, searching for a local enhancement of the cross-section around some value of \mjj, which is what ``resonance'' means in this context.  The comparison is made by a hypertest known as {\sc BumpHunter} \cite{BH}.  Bayesian lower limits are set, at 95\% credibility level (C.L.), on the mass of excited quarks \cite{exq}, axigluons \cite{axg}, and scalar color octets \cite{TaoHan}.  Finally, model-independent limits are given, which can be used to approximate the limit to virtually any resonant signal decaying into two jets.

\begin{figure}
\subfigure[]{
  \includegraphics[width=0.45\textwidth]{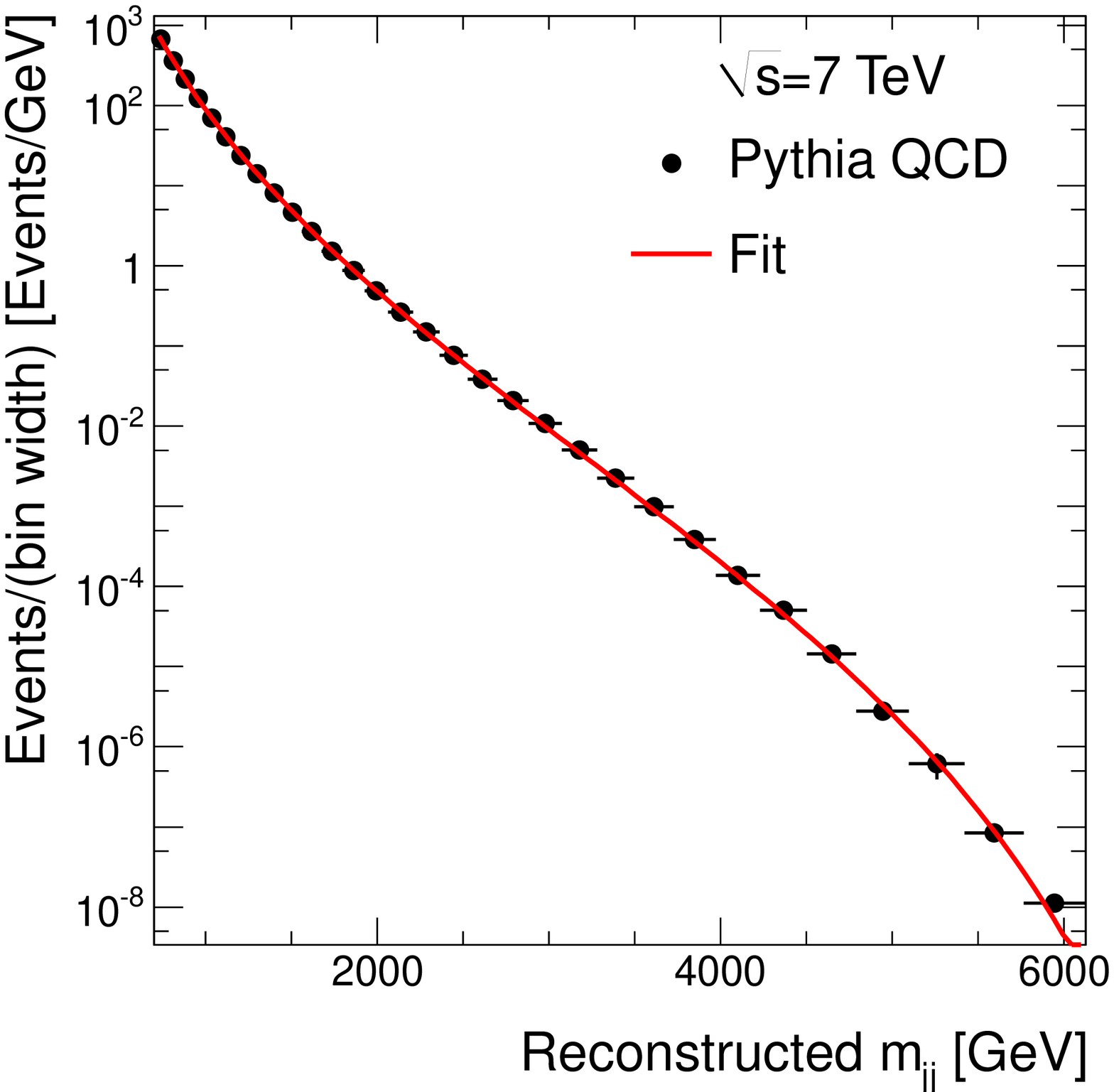}
\label{fig:fitToPythia}
}
\subfigure[]{
  \includegraphics[width=0.45\textwidth]{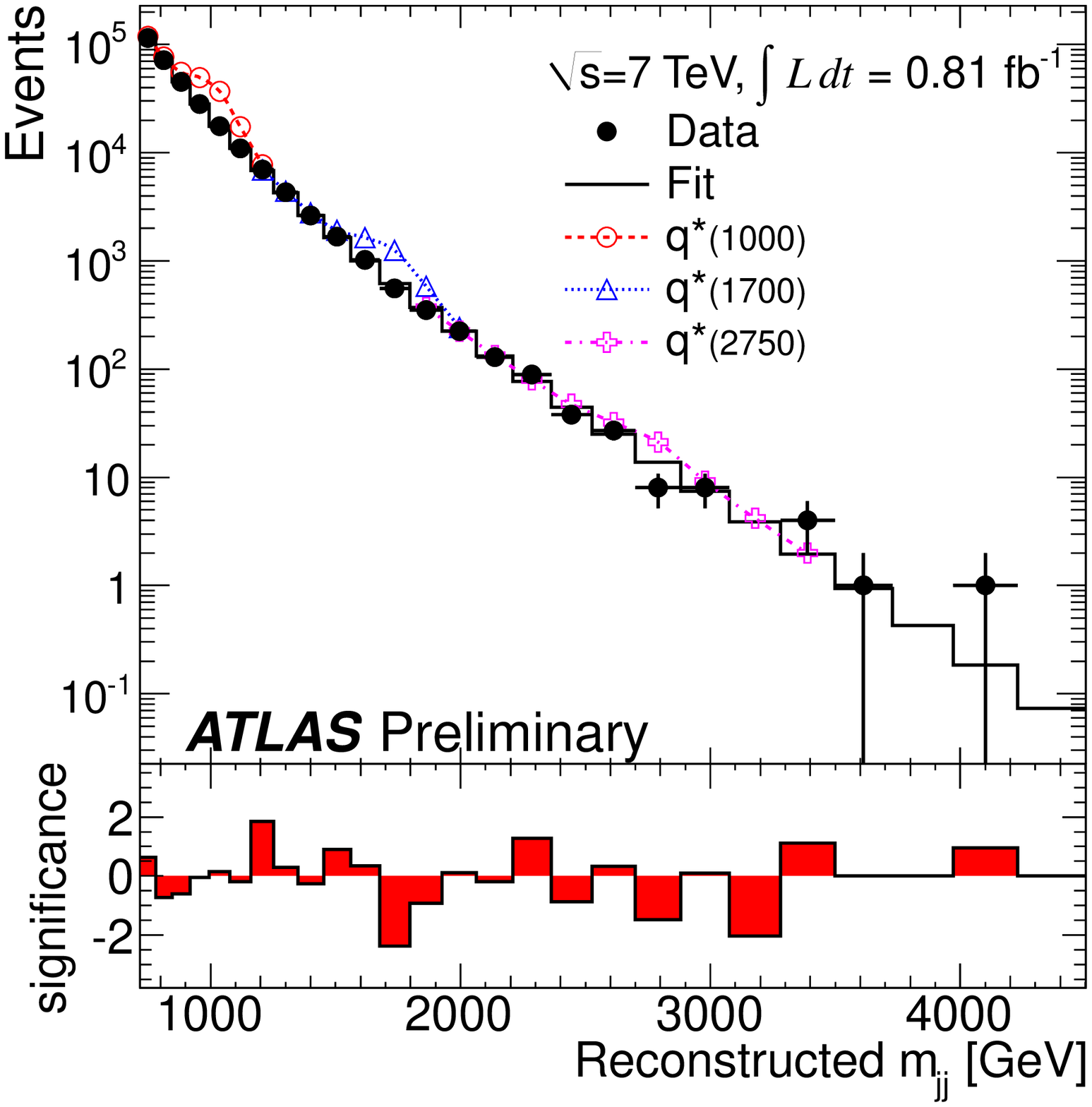}
\label{fig:mjj}
}
\caption{Left: The function of eq.~\ref{eq:function} used to fit {\sc Pythia} QCD, after ATLAS detector simulation.  Right: The data (black markers), the fitted background (solid histogram), and three examples of $q^\ast$ signal (hollow markers).  The red bars indicate the significance of each excess (positive bars) or deficit (negative bars) of data, expressed in standard deviations \cite{PDG}.  
In bins with small expected number of events, where the observed number of events is similar
to the expected, the Poisson probability of a fluctuation at least as high (low) as the observed
excess (deficit) can be greater than 50\%, resulting in a negative significance when expressed in standard deviations.  Drawing negative values would obscure the current intuitive convention that data excesses appear as positive bars and data deficits as negative bars, and the amplitude of each bar is proportional to the significance of each deviation.  
Such bins present no statistical interest, so, for simplicity, bars are not drawn for them.}
\end{figure}

The expected spectrum is obtained by fitting to the data the function
\begin{equation}
f(x) = p_0\frac{(1-x)^{p_1}}{x^{p_2 + p_3\ln x}},\hspace{0.15cm}  \text{where\ \ } x\equiv \frac{\mjj}{\sqrt{s}}.
\label{eq:function}
\end{equation}
This parametrization has been shown to fit well the QCD prediction given by {\sc Pythia}, {\sc Herwig}, {\sc Alpgen}, and {\sc NLOJET++} \cite{CDFanalysis}.  Figure~\ref{fig:fitToPythia} shows an example of fitting the {\sc Pythia} QCD prediction, after full ATLAS  detector simulation, without NLO corrections.

The fit is performed in such a way that, if the data contains non-negligible signal in any \mjj interval, that will be omitted from the fit, thus obtaining the background from the sidebands.  The algorithm that determines if that is necessary searches for any \mjj window that may be responsible for a globally poor fit ($\chi^2$ test \pval $<$ 1\%).  No such interval was found in the data, so, the whole spectrum was fitted, giving a $\chi^2$ \pval much greater than the 1\% threshold.

Figure \ref{fig:mjj} shows the data, along with the fitted background, and three examples of $q^\ast$ signal added to it.  The overall consistency of the data with the fit is quantified by the \pval of the negative logarithm of the likelihood of the data when the background is expected ($-\log L(\text{data}|\text{background})$).  This is a generalization of the $\chi^2$ test that accounts correctly the Poisson probability in bins with low background.  That \pval is about \pvalL\%, indicating consistency.

\subsection{The model-independent search for resonances}
\label{sec:searchPhase}

The {\sc BumpHunter} algorithm \cite{BH} is used to look for resonances in the \mjj spectrum in a model-independent way, which offers high sensitivity, and takes correctly into account the trials factor (a.k.a. ``look elsewhere'' effect), namely the effect of examining various positions of the \mjj spectrum before eventually finding something.
The general way to account for the trials factor is to construct a hypertest, and the {\sc BumpHunter} is just one such hypertest.

In \cite{BH} one can find a detailed account of the trials factor, the definition of hypertests, and details about the {\sc BumpHunter} and other similar hypertests, such as the {\sc TailHunter}.  A summary is given here.

A hypertest is a hypothesis test, much like the well-known $\chi^2$ or the Kolmogorov-Smirnov (KS) test.  It has a test statistic, and a corresponding \pval, which is, as usual, the frequency with which a more signal-like test statistic than the one observed in data would occur under the background-only hypothesis.  The difference is that, in a hypertest, the test statistic {\em itself} is a \pval\footnote{As we will see, it's actually a monotonically decreasing function of a \pval, like $(-\log{\pval})$, for reasons of convention.}, therefore the \pval of a hypertest is a \pval of a \pval.\footnote{Contrast that to the $\chi^2$ test or the KS test, where the test statistic is just a metric of discrepancy, like the $\chi^2$ or the biggest difference between two cumulative distributions.}  Each hypertest contains in its definition an ensemble of hypothesis tests.  These are the hypertest's ``members'', and their multitude is responsible for the trials factor.  The hypertest combines the results of its members\footnote{A hypertest can contain even {\em hyper}tests; nothing changes.   Combining simple hypothesis tests creates a hypertest.  Combining hypertests creates just another hypertest.  A hypertest that contains just one member test is a trivial hypertest, whose \pval is identical to the \pval of its member.}.  The hypertest's test statistic is, by convention\footnote{The point of this convention is to have a test statistic which {\em increases} with increasing discrepancy.}, the negative logarithm of the smallest \pval of all tests in the ensemble.  The \pval of the hypertest is the frequency by which, in the background-only hypothesis, there would be \emph {any} test in the ensemble with a \pval at least as small as the smallest \pval found in the ensemble of tests when they were performed on the actual data.

The {\sc BumpHunter}, as implemented in this analysis, is a hypertest whose ensemble of combined tests is defined after forming all possible \mjj intervals in the binned \mjj spectrum, and performing in each interval a simple event-counting hypothesis test.   The bin sizes increase at higher \mjj, following detector resolution, such that any new physics signal would populate at least two consecutive bins.  For this reason, we do not consider \mjj intervals narrower than two bins.  Since we are only interested in {\em excesses} of data, the test statistic of the hypothesis test performed in each \mjj interval takes its minimum value, which is 0 by convention, if the data in the interval ($D$) are fewer than expected ($B$).   If $D > B$, then the test statistic increases monotonically with $D-B$, for example as $(D-B)^2$, or $(D-B)^{100}$, or $-\log(\frac{B^D}{D!}e^{-B})$; it doesn't matter exactly how the test statistic increases\footnote{It doesn't even matter if the test statistic is $(D-B)^2$ in one interval and $(D-B)^{100}$ in another.}, because all the hypertest needs is the \pval of this test statistic, which in any case is the Poisson probability of observing at least $D$ events when $B$ are expected: $\sum_{n=D}^{\infty} \frac{B^D}{D!}e^{-B}$.

An equivalent, more procedural (and maybe more intuitive) way to describe the {\sc BumpHunter} hypertest is the following algorithm:
\begin{enumerate}
\item Scan the \mjj spectrum, counting events in all intervals of at least 2 bins, and in each interval compute the Poisson \pval of seeing at least as many data as observed, given the expected number of events.
\item Keep the smallest Poisson \pval found in the actual data.   This number is the {\em observed} test statistic of the {\sc BumpHunter}. \label{step2}
\item Generate many spectra of background-only pseudo-data, and scan them too, noting the smallest Poisson \pval in each one.  Note that, since the data were compared to the background fitted to them, the fit must be repeated to each individual pseudo-spectrum before scanning.  This wouldn't be necessary if the background was not obtained by a fit.
\item Measure how frequently a pseudo-spectrum returns a minimum Poisson \pval that is smaller than the observed test statistic (step \ref{step2}).  This frequency is the \pval of the {\sc BumpHunter} hypertest.
\end{enumerate}

If the {\sc BumpHunter}'s \pval is very small\footnote{By pure convention, a \pval of $1.35\times 10^{-3}$, corresponding to a 3$\sigma$ effect \cite{PDG}, is considered ``evidence'', and a \pval of $2.87\times 10^{-7}$, corresponding to a 5$\sigma$ effect, is considered ``discovery''.  One may argue that the 5$\sigma$ requirement is unnecessarily conservative, especially for the \pval of a hypertest where the trials factor is already accounted for.} it is a fail-safe statement to say that there is a local excess in the data, which is incompatible with the background, because the probability of such an excess being pure coincidence is equally small.  It also helps that we know which member test (i.e.\ which \mjj interval) returned the smallest \pval, because that is where the signal is, so we are pointed at it.\footnote{However, it should not be thought that the {\sc BumpHunter} is performing an inference of the actual mass or width of a new particle.  It is just a frequentist test of the background-only hypothesis, which returns a \pval; not any confidence interval or posterior probability density for the signal parameters.  The interval pointed at by the {\sc BumpHunter} could be used as input to a formal inference procedure, which may be Bayesian or Frequentist, and which will have to assume some more information about the signal, e.g.\ the shape of its distribution in \mjj.}

By its very construction, a hypertest accounts for the trials factor, since it considers every member test in every pseudo-experiment.  

It is very important that the hypertest's test statistic is the largest $-\log(\pval)$ in the ensemble, e.g.\ the largest
\begin{equation}
  -\log\left(\sum_{n=D}^{\infty} \frac{B^n}{n!}e^{-B}\right),
\end{equation}
and {\em not} the largest test statistic in the ensemble, e.g.\ the largest
\begin{equation}
-\log\left(\frac{B^D}{D!}e^{-B}\right),\mbox{ or } (D-B)^{100}.
\end{equation}
It would have been wrong to use directly the test statistic of the most discrepant member test, because test statistics of different member tests follow different distributions under the background-only hypothesis, and are not comparable.  For example, in the case of the {\sc BumpHunter}, this mistake could have led to a hypertest with strong bias towards low \mjj regions with high background ($B$), because, e.g., $(1010 - 1000)^{100}$ is greater than $(15-10)^{100}$, even though it's obvious that observing 1010 events instead of 1000 is less significant than observing 15 instead of 10.  The right way to combine dissimilar hypothesis tests is by comparing their \pvals, not their test statistics.  Hypertests, by construction, avoid this pitfall.

\begin{figure}
\subfigure[Toy background and signal] {
  \includegraphics[width=0.45\textwidth]{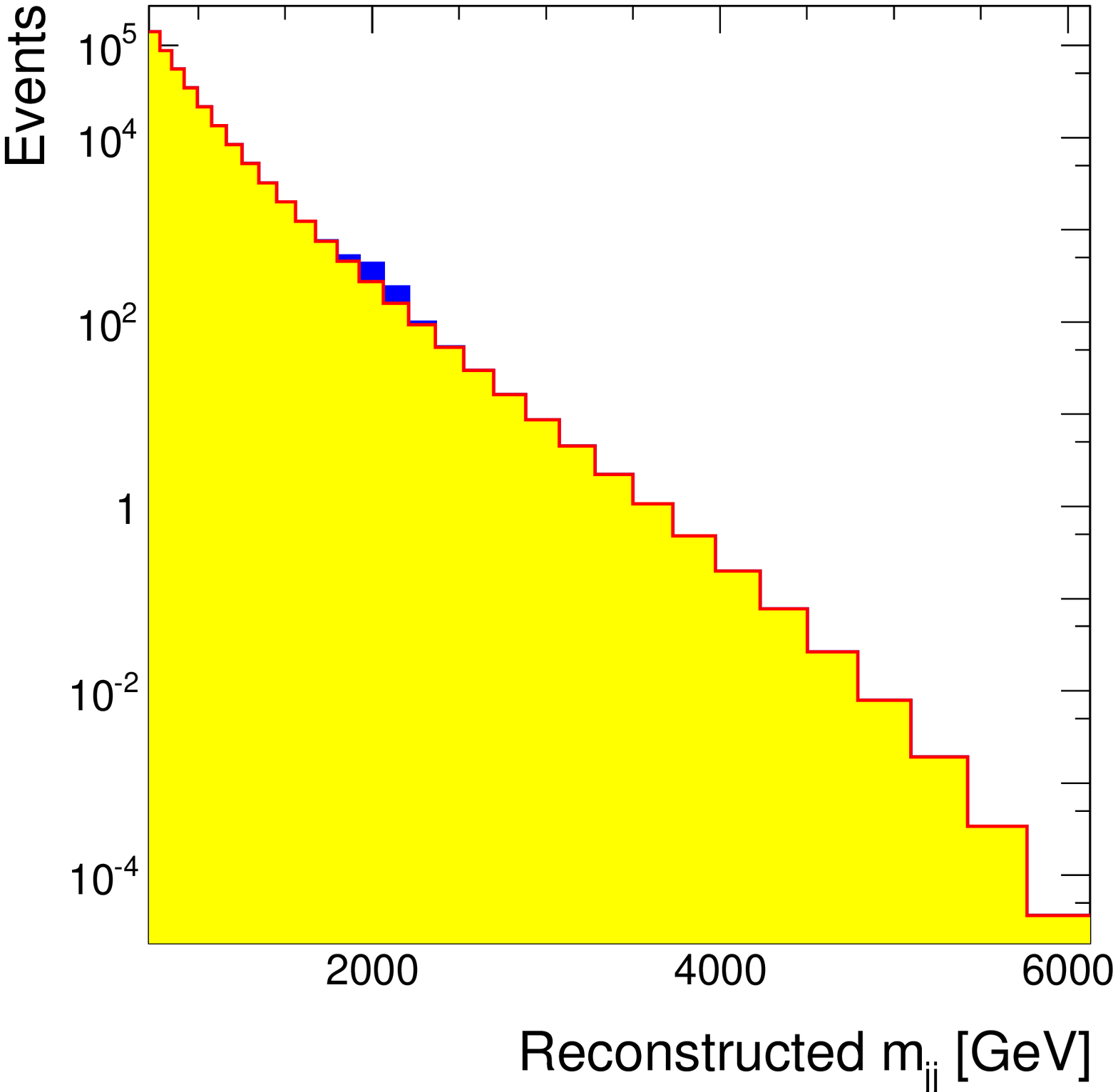}
  \label{fig:sensSignal}
}
\subfigure[Power of tests] {
  \includegraphics[width=0.45\textwidth]{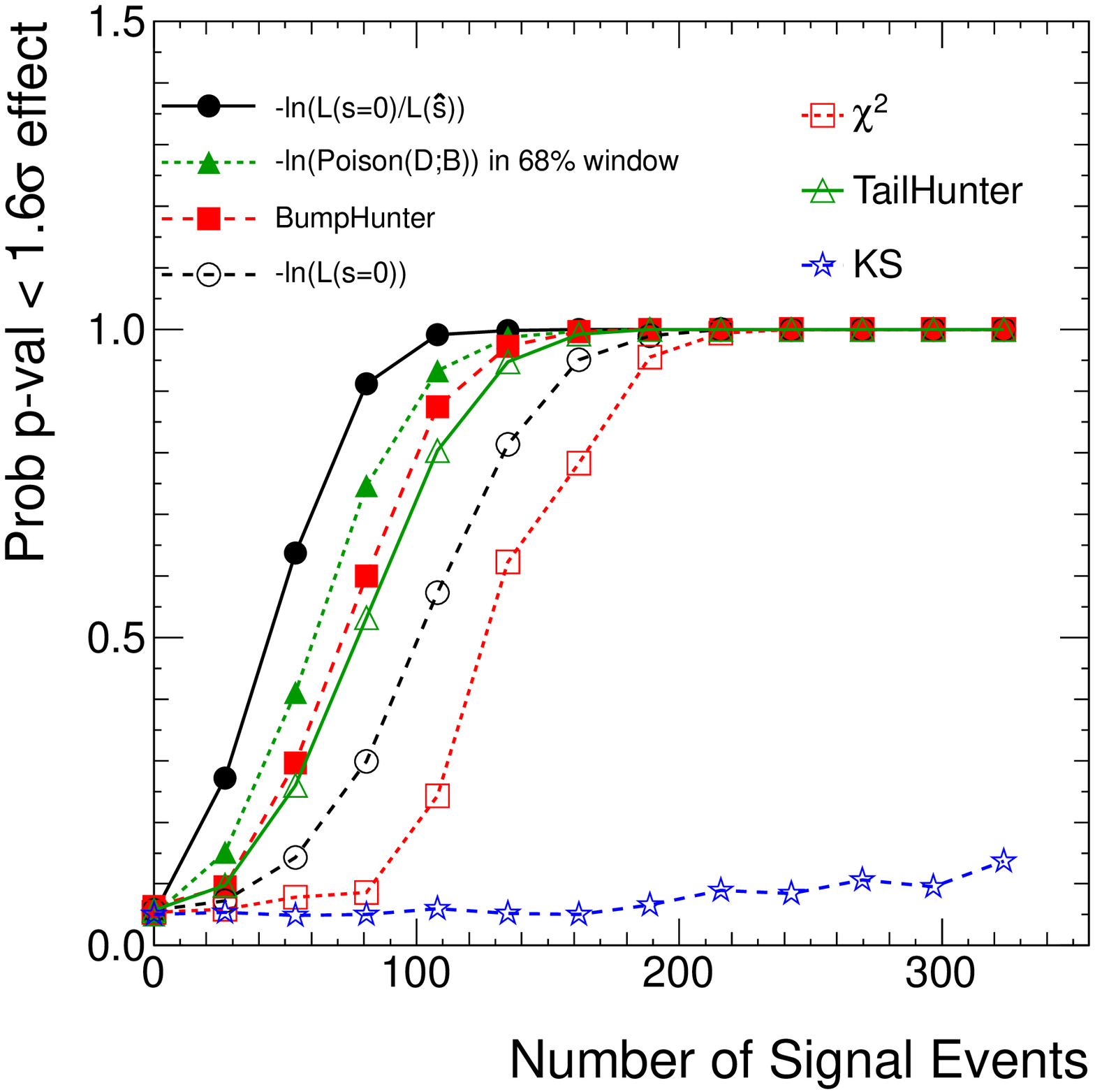}
  \label{fig:sensPower}
}
\caption{Left: The toy signal (blue, Gaussian at 2 TeV with $\sigma$=100~GeV) injected on top of a background similar to QCD.
  Right:  The power (see text) of the following tests, as a function of injected signal events:
  The {\sc BumpHunter} (filled squares), the {\sc TailHunter} \cite{BH} (empty triangles), the KS test (stars), the $\chi^2$ test (empty squares), its generalization that uses the negative logarithm of the likelihood of the data under the background-only hypothesis (empty circles), and two tests that know something about the injected signal:  a test that performs event-counting in the windw which contains 68\% of the signal (filled triangles), and a test that knows not only the position, but also the exact shape of the signal, and computes a likelhood ratio test statistic (filled circles). \label{fig:sensitivity}}
\end{figure}

Figure~\ref{fig:sensitivity} demonstrates the sensitivity of the {\sc BumpHunter} to a test signal injected at 2~TeV, with Gaussian shape of standard deviation 100~GeV (Fig.~\ref{fig:sensSignal}).   Figure~\ref{fig:sensPower} compares the power of a variety of hypothesis tests, defined as the probability a test has to observe a \pval less than 5\% (equivalent to a 1.6$\sigma$ effect)\footnote{This is an arbitrary choice that doesn't affect the conclusions of this comparison.}, as a function of the expected number of injected signal events.  The faster the power increases, the higher the sensitivity of a test to this signal.

In Fig.~\ref{fig:sensPower}, two of the compared tests, the {\sc BumpHunter} and the {\sc TailHunter} \cite{BH}, are hypertests with no knowledge of the injected signal.  The rest are not hypertests.  Three of them examine the whole spectrum: the KS test, the $\chi^2$ test, and its generalization that uses the test statistic $-\log(\prod_{\text{bins}} \frac{B^D}{D!}e^{-B})$.  Finally, two hypothesis tests are constructed exploiting knowledge of the signal, which in reality is not possible, unless one knows in advance what is about to be discovered.  One of these two tests is aware only of the signal location, so, it performs a counting test just in the bins where 68\% of the signal is expected.  The other knows the exact shape and position of the signal, but not how much it is, so it uses as test statistic the $\log\left(\frac{L(\text{data}|\hat{s})}{L(\text{data}|\text{no signal})}\right)$, where $\hat{s}$ is the best fitting amount of the known signal.

The comparison shows that, among the tests with no prior knowledge of the signal, the {\sc BumpHunter} is by far the most sensitive, followed by the {\sc TailHunter}.  This is true despite the trials factor, which does not apply to other tests.  
The KS test is the least sensitive to this signal.  The most sensitive test of all is the one which knows the exact shape and position of the signal.  The next most sensitive is the one that knows its position.  

\subsubsection{Results of the search for resonances}

\begin{figure}
\subfigure[Most discrepant interval] {
\includegraphics[width=0.31\textwidth]{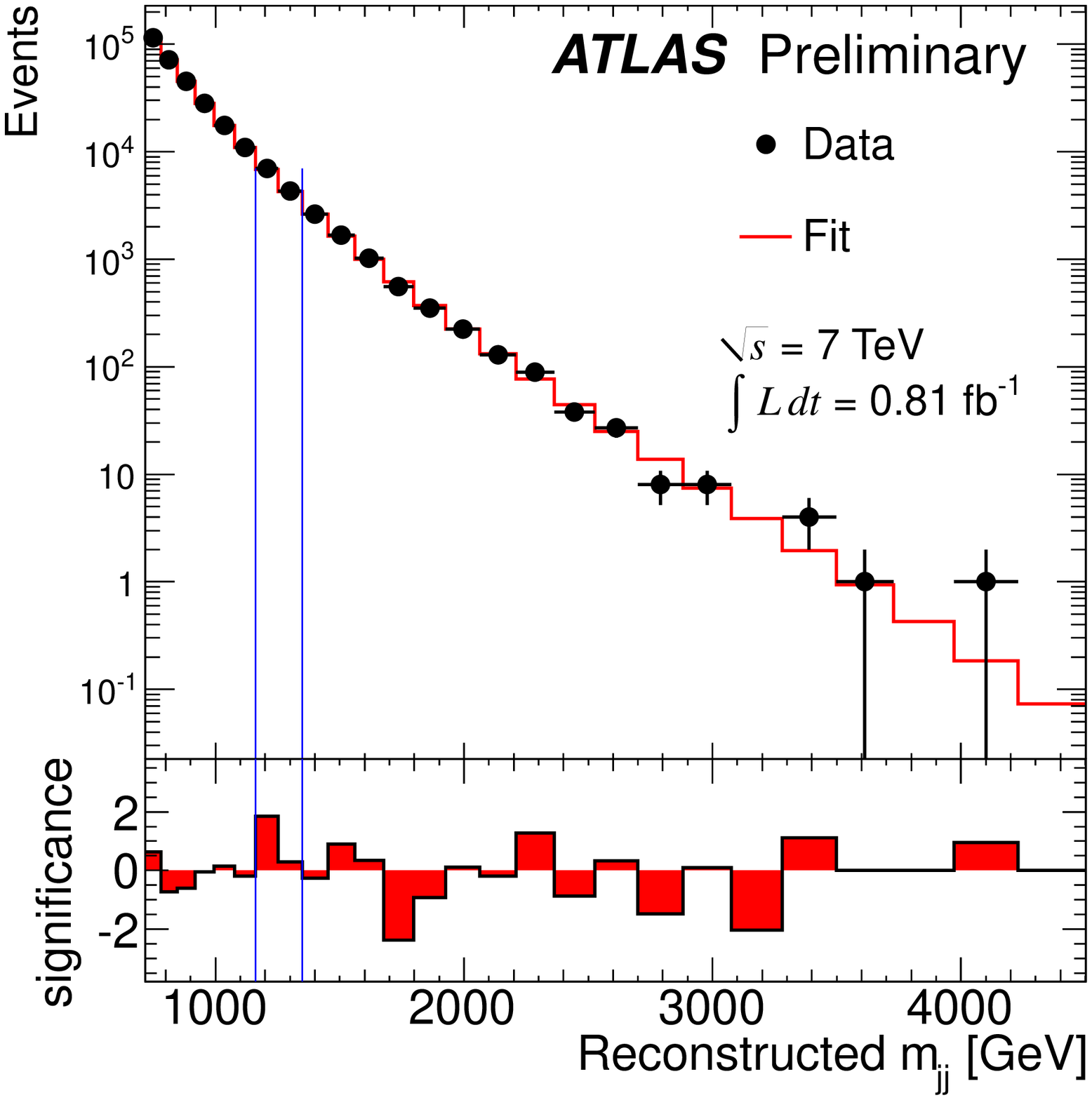}
\label{fig:mostInteresting}
}
\subfigure[{\sc BumpHunter} tomography plot] {
\includegraphics[width=0.31\textwidth]{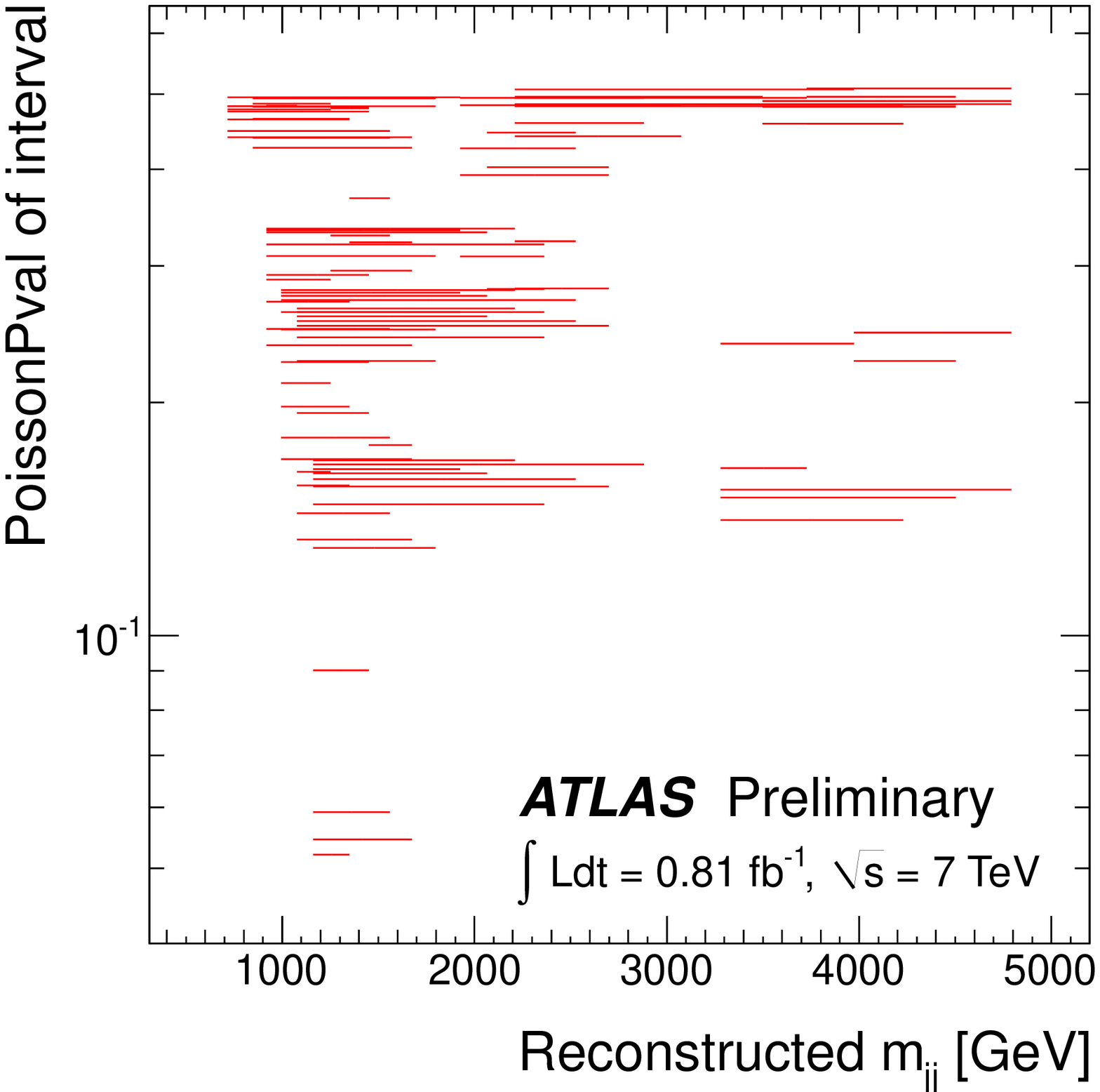}
\label{fig:tomography}
}
\subfigure[Distribution of test statistic] {
\includegraphics[width=0.31\textwidth]{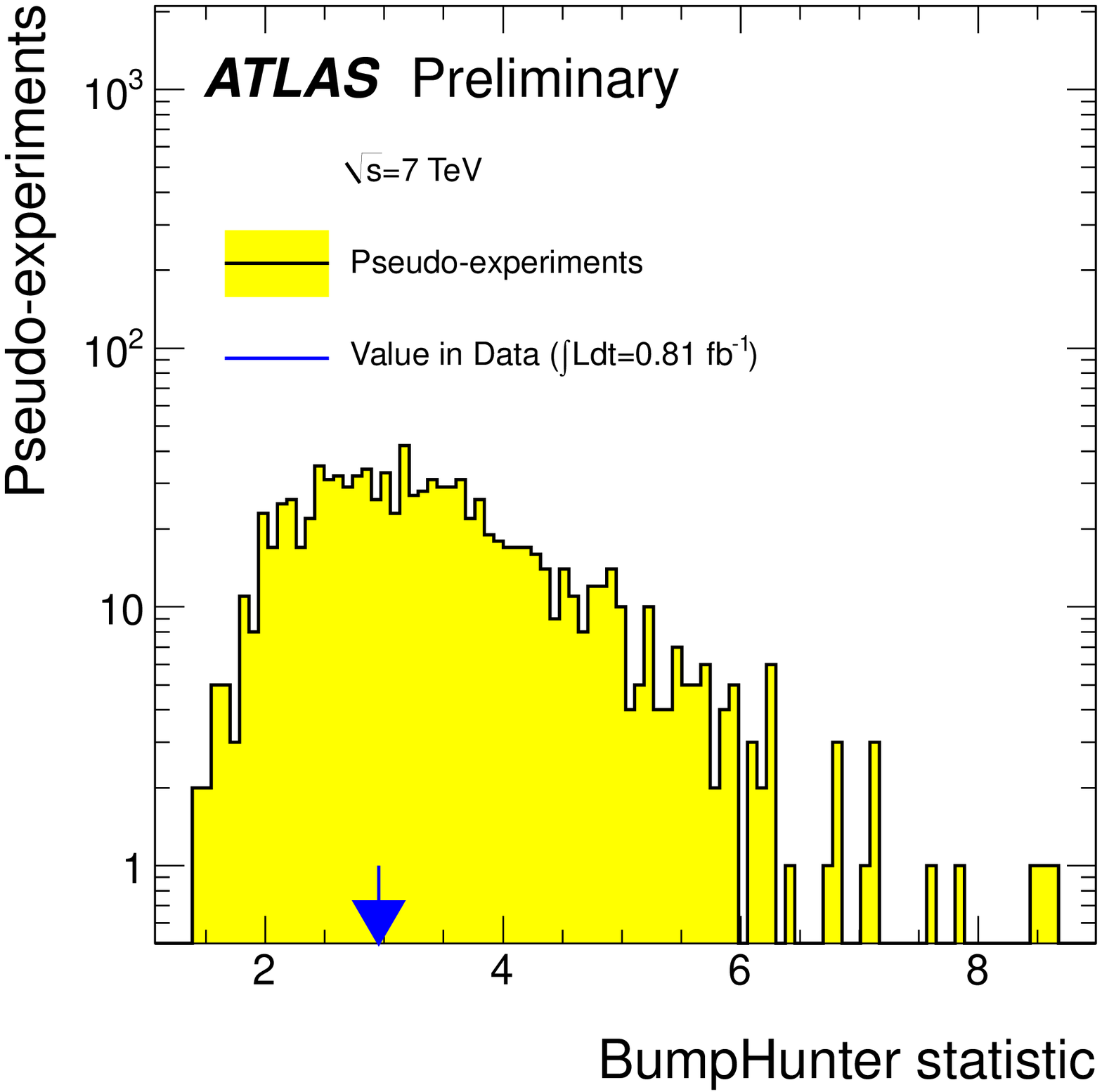}
\label{fig:nullStatistic}
}
\caption{Results of the search for resonances. 
Left: The most significant excess found in the data, indicated by the blue vertical lines. 
Middle: The Poisson \pval ($\sum_{n=D}^{\infty} \frac{B^n}{n!}e^{-B}$ in each interval with an excess ($D>B$).  Each \mjj interval is indicated by its position in the horizontal axis, and its \pval by the position on the vertical axis.  Even the most significant interval, without any trials factor to be considered, has a \pval of about 0.09, which is very insignificant. 
Right: The distribution of the {\sc BumpHunter} test statistic under the background-only hypothesis, compared to the observed test statistic (blue arrow).  The {\sc BumpHunter} \pval is \bhpval\%.
\label{fig:resultsOfSearch}}
\end{figure}

Figure~\ref{fig:mostInteresting} shows the most interesting interval identified by the {\sc BumpHunter} in the actual data.  
Figure~\ref{fig:tomography}, known as ``{\sc BumpHunter} tomography'', shows $\sum_{n=D}^{\infty} \frac{B^n}{n!}e^{-B}$ in each \mjj interval that contains an excess of data ($D>B$), which allows us to see that there is no interval that gets anywhere close to being significant, even without accounting for the trials factor.
Figure~\ref{fig:nullStatistic} shows the distribution of the {\sc BumpHunter} test statistic under the background-only hypothesis, and the blue arrow indicates the observed test statistic value.  From that it is clear that the observed spectrum does not contain any significant bump anywhere.  The {\sc BumpHunter}'s \pval is \bhpval\%.

No significant excess has been found in \mjj in \lumi of data.

\subsection{Limits to specific models}
\label{sec:limits}

\begin{figure}
\subfigure[Limits to $q^\ast$ and $A$] {
\includegraphics[width=0.31\textwidth]{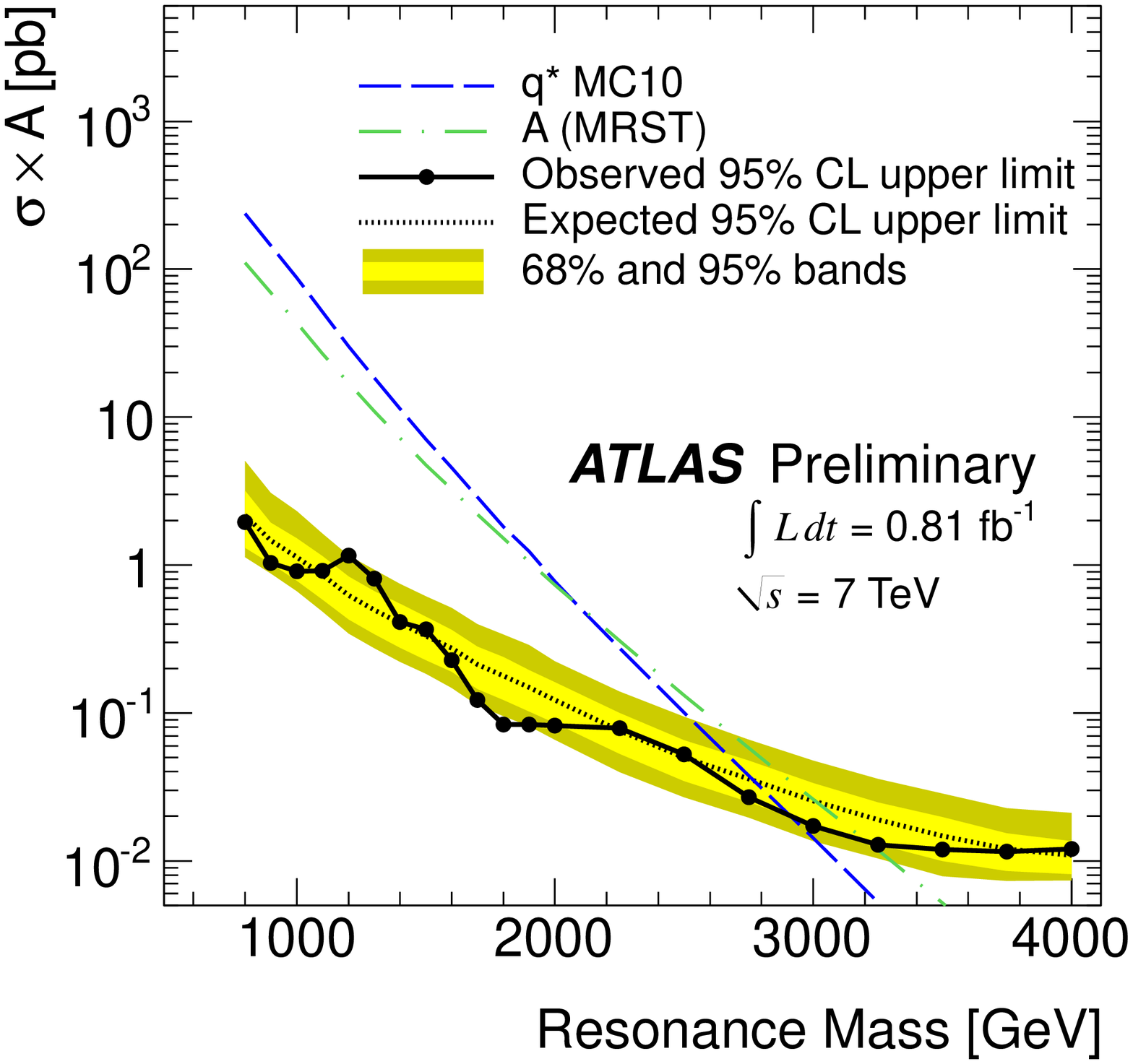}
}
\subfigure[Limits to $q^\ast$ and $A$] {
\includegraphics[width=0.31\textwidth]{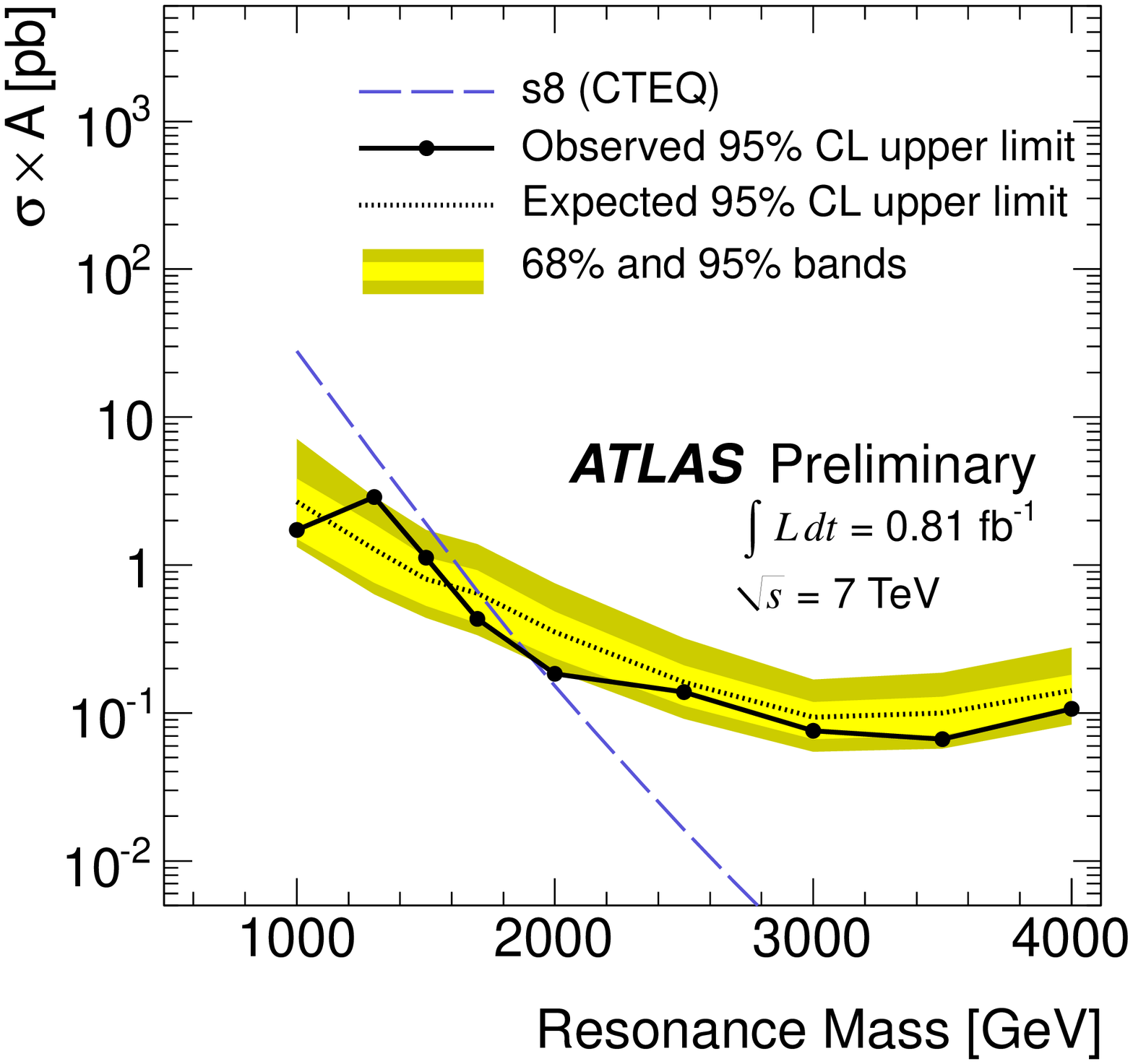}
}
\subfigure[Limits to Gaussians] {
  \includegraphics[width=0.31\textwidth]{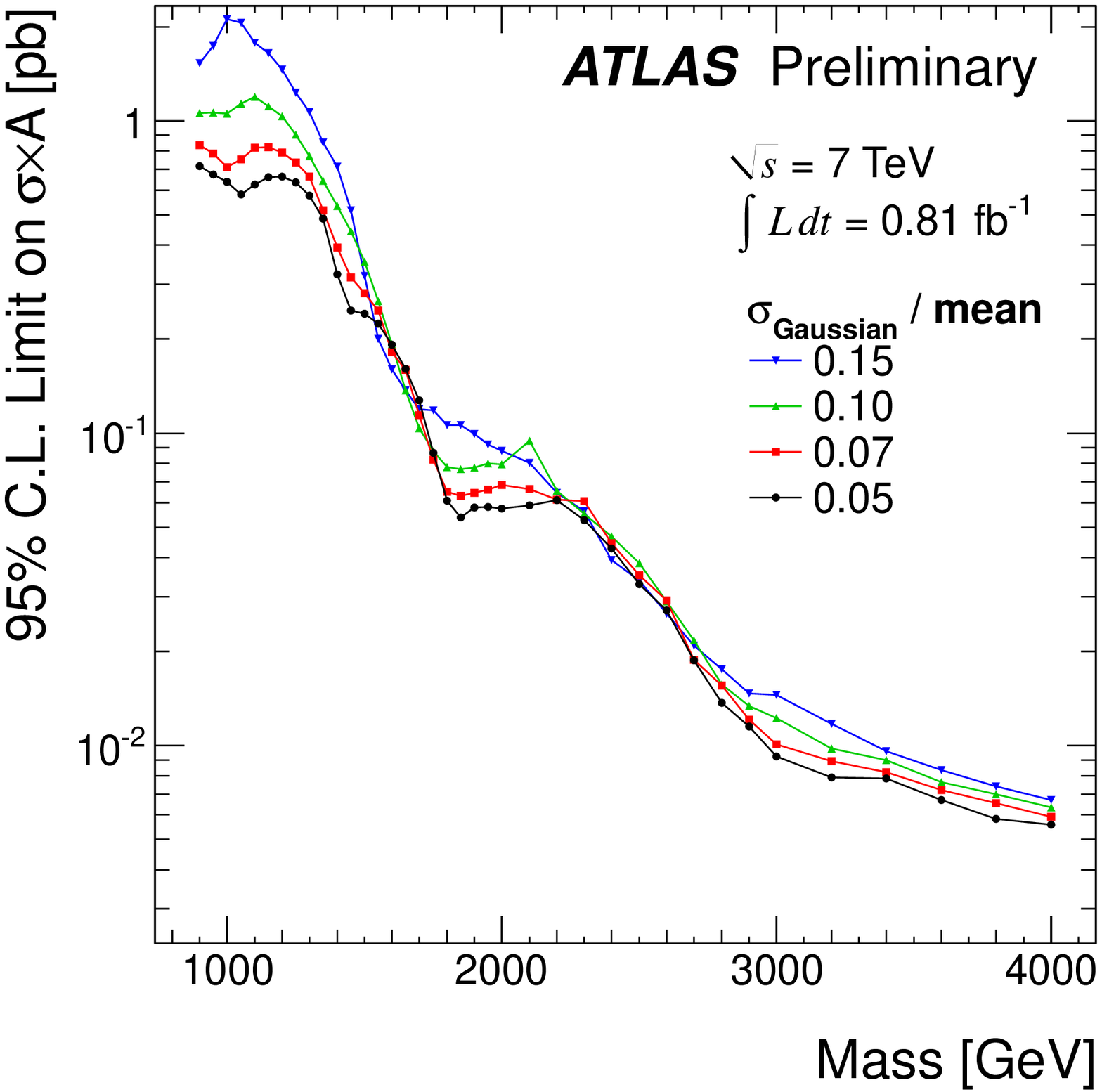}
  \label{fig:tabular}
}
\caption{Limits to excited quarks and axigluons (left), to scalar color octets (middle), and to Gaussian-distributed signals (right) of mean values between 900 GeV and 4 TeV and standard deviations ($\sigma_{\rm Gaussian}$) from 5\% to 15\% of the mean value.
\label{fig:limits}}
\end{figure}

Upper limits are set to the accepted cross-section ($\sigma \times \mathcal{A}$) of excited quarks ($q^\ast$), axigluons ($A$), and scalar color octets ($s8$).  Details about how these models were simulated can be found in \cite{resonancePLB}.  The limits are Bayesian, at 95\% credibility level, with a constant prior in $\sigma \times \mathcal{A}$.  The observed and expected limits are compared to the corresponding theoretical predictions in Fig.~\ref{fig:limits}.  Lower mass limits are summarized in Table~\ref{table:limitsResonance}.

The main systematic uncertainties, which have been convolved to the limits, are the jet energy scale (JES) uncertainty (2 to 4\%, depending on \pt and $\eta$), the background fit uncertainty (ranging from less than 1\% at the beginning of the spectrum, and reaching about 20\% at 4~TeV), and the luminosity uncertainty of \lumUncert\%.  The jet energy resolution (JER) uncertainty has a negligible effect on the limits and is ignored.

\begin{table}
\caption {The 95\% CL mass lower limits for the models 
examined in this study.}
\label{table:limitsResonance}
\begin{tabular}{lcc}
\hline
Model \quad \quad &  \multicolumn{2}{c}{95\%\ CL Limits (TeV)} \\
            &  Expected  &  Observed\\
\hline 
   Excited Quark  &   \limExqExp   &  \limExqObs  \\ 
   Axigluon                &   \limAExp &  \limAObs \\  
   Colour Octet Scalar     &   \limSExp  &  \limSObs  \\  
\hline
\end{tabular}
\end{table}

\subsection{Model-independent limits}
\label{sec:MIlimits}

In addition to specific theories, limits are also set to a collection of hypothetical signals that are assumed to be Gaussian-distributed in \mjj, with means ranging from 0.9 to 4.0 TeV, and standard deviations from 5\% to 15\% of the mean.  These limits include the same luminosity and background fit systematic uncertainties.  Since the Gaussian \mjj distributions do not result from actual jets, to convolve the JES uncertainty the mean of the Gaussian is given a 4\% uncertainty, which is conservatively larger than the shift observed in $q^\ast$ templates when jet \pt was shifted by the JES uncertainty.

The results are shown in Fig.~\ref{fig:tabular}.   These limits can be used to set approximate limits on any new physics model that predicts some peaking signal in \mjj, because almost every signal excess can be approximated by a Gaussian at some level.   A procedure is given in \cite{resonancePLB}, to make the approximation.   The non-Gaussian tails of the signal need to be removed, to make the Gaussian shape a better approximation to the remaining signal.  Removing the tails results in lower signal acceptance.  Removing the tails does not affect the much the mass limits, because the tails are in background-dominated regions; the theoretical cross-section of the new particle is reduced by the acceptance of keeping just the core of its \mjj distribution, but the $\sigma \times \mathcal{A}$ upper limit is reduced by roughly the same fraction, so, the theory and the limits continue to intersect at the same mass as if the tails had not been removed.

\section{Conclusion}

The latest results were shown, from the new physics searches in the dijet angular and mass distribution.  The latter was updated with \lumi of 2011 data.  No sign of new physics was found.  Limits to $q^\ast$ improved by about 1 TeV since 2010, and are the most stringent ones currently.  Limits to Gaussian signal distributions have been updated, as a tool to compute approximate limits to more theoretical models.

\begin{acknowledgments}
The author thanks the organizers of DPF2011 and the National Science Foundation for their financial aid, and for the opportunity to share these results.
\end{acknowledgments}

\bigskip 

\end{document}